\documentclass[transaction]{IEEEtran}
\usepackage{times}
\usepackage{fancyhdr,graphicx,amsmath,amssymb}
\usepackage[ruled,vlined]{algorithm2e}
\usepackage{graphicx}
\usepackage{enumitem}
\usepackage{subcaption}
\usepackage{multirow}
\usepackage{mathtools}
\usepackage{balance}
\usepackage{url}
\usepackage{booktabs}
\usepackage{breqn}
\usepackage{amsfonts,amssymb}
\usepackage[noadjust]{cite}
\usepackage{color}

\usepackage[]{algorithm2e}

\begin{document}

\title{FMNet: Latent Feature-wise Mapping Network for Cleaning up Noisy Micro-Doppler Spectrogram}

\author{
    \IEEEauthorblockN{Chong Tang\IEEEauthorrefmark{1}, Wenda Li\IEEEauthorrefmark{1}, Shelly Vishwakarma\IEEEauthorrefmark{1}, Fangzhan Shi\IEEEauthorrefmark{1}, Simon Julier\IEEEauthorrefmark{2}, Kevin Chetty\IEEEauthorrefmark{1}}
    \IEEEauthorblockA{\\\IEEEauthorrefmark{1}Department of Security and Crime Science, University College London, UK}\\
    \IEEEauthorblockA{\IEEEauthorrefmark{2}Department of Computer Science, University College London, UK}
    \\{\IEEEauthorrefmark{1}{{chong.tang.18, s.vishwakarma, wenda.li, fangzhan.shi.17, k.chetty}}@ucl.ac.uk,\IEEEauthorrefmark{2}{s.julier}@ucl.ac.uk}
}

\maketitle

\begin{abstract}
Micro-Doppler signatures contain considerable information about target dynamics. However, the radar sensing systems are easily affected by noisy surroundings, resulting in uninterpretable motion patterns on the micro-Doppler spectrogram. Meanwhile, radar returns often suffer from multipath, clutter and interference. These issues lead to difficulty in, for example motion feature extraction, activity classification using micro Doppler signatures ($\mu$-DS), etc. In this paper, we propose a latent feature-wise mapping strategy, called Feature Mapping Network (FMNet), to transform measured spectrograms so that they more closely resemble the output from a simulation under the same conditions. Based on measured spectrogram and the matched simulated data, our framework contains three parts: an Encoder which is used to extract latent representations/features, a Decoder outputs reconstructed spectrogram according to the latent features, and a Discriminator minimizes the distance of latent features of measured and simulated data. We demonstrate the FMNet with six activities data and two experimental scenarios, and final results show strong enhanced patterns and can keep actual motion information to the greatest extent. On the other hand, we also propose a novel idea which trains a classifier with only simulated data and predicts new measured samples after cleaning them up with the FMNet. From final classification results, we can see significant improvements.
\end{abstract}

\begin{IEEEkeywords}
Micro-Doppler Spectrogram, Adversarial Autoencoder, Variational Autoencoder, Feature Mapping, Passive WiFi Radar, Deep Learning, Activity Classification
\end{IEEEkeywords}

\IEEEpeerreviewmaketitle

\section{Introduction}
\label{Introduction}
Micro-Doppler spectrograms ($\mu$-DS) express target motion and micro-motion characteristics in the time-frequency domain and has been extensively used in many applications; from people counting and human motion classification, to hand gesture and drone recognition\cite{tang2020occupancy, li2018passive, gurbuz2020radar, palama2021measurements}. In practice, the quality of $\mu$-DS is easily affected by background clutter and multipath reflections, which leads to weak Doppler strength and blur motion patterns. Recently, solutions based on deep learning (DL) have attracted a lot of research interest. These methods normally require clean spectrograms as labels to train a neural network, but practically, it is hard to collect clean measured data. To provide sufficient training data, methods like\cite{huang2018micro, rock2019complex} manually added Additive White Gaussian Noise (AWGN) to simulation spectrograms to obtain enough training pairs. However, the disadvantage to this approach is that such a artificial noise is pixel-independent and rarely matches the spatially correlated real-world noise. The study in \cite{tang2021learning} uses close-to-nature noise produced by a Generative Adversarial Network (GAN) to replace AWGN, which improved the denoising performance. But, another issue is that both noise modelling methods cannot effectively synthesize multipath effects. For the indoor sensing scenarios which can be significantly affected by multipath clutter, these methods only play limited roles. In this case, it is necessary to consider a more efficient and robust alternative. 

Whether traditional or DL-based algorithms, they serve the same purpose: to generate the clean motion information. In this paper, we describe the purpose as the $\mu$-DS enhancement task. Among them, DL methods are more flexible and easier to implement. We can combine it with various existing radar sensing systems without modifying the underlying signal processing workflow. Meanwhile, the feature extraction ability of convolutional operations \cite{rawat2017deep} in DL can learn complex spatial-temporal correlations of $\mu$-DS. These advantages make it worth further exploring a better enhancement strategy using DL algorithms. On the other hand, the aforementioned DL-based studies expended many effort in creating training pairs by artificially adding noise. But again, this is not always feasible. Some phenomena, such as multipath clutter, are unique in each case and cannot be approximated in that manner. So, we should approach this issue from a different perspective.

While environmental factors and multipath have an impact on measured $\mu$-DS, motion information remains the most significant component. Du et al.~\cite{du2015noise} proposed a Beta process-based principle component analysis (PCA) to reduced interference by reconstructing radar returns from only principle motion components in the subspace. Other studies like~\cite{saho2017gait, du2014noise, an2020range} also applied PCA-based method to reduce data dimensionality to enhance $\mu$-D motion information. Such methods of directly analysing the major motion information in low-dimensional space shows a more effective way of thinking. In DL structures, an alternative for performing similar functions is Autoencoder (AE)\cite{hinton1994autoencoders}. The network has an Encoder-Decoder structure, in which the encoder extracts latent features from the input and the decoder reconstructs the output using the latent features. Herein, the latent features form a vector which is another representation of a spectrogram. By minimizing the reconstruction loss, the net can automatically keep the desired features in the latent vector and discard other noise. The AE-based framework has been extensively applied in fields like the speech enhancement\cite{araki2015exploring, xia2013speech} and image denoising\cite{gondara2016medical, vincent2008extracting, nishio2017convolutional}. The advantage of the AE is that it can efficiently accomplish dimensionality reduction and reconstruct data with a single network. Meanwhile, the convolutional operations endow AE with superior feature extraction capabilities for the image-like $\mu$-DS data.

Based on the AE's characteristics, we propose a novel $\mu$-DS enhancement framework termed FMNet. Unlike DL-based approaches, the FMNet is based on the dimensionality reduction abilities of AE, and bypasses unimportant information to directly enhance motion patterns based on the similarity of latent features between measured and simulated data. This kind of tasks can be achieved by applying the adversarial structure which is able to mapping one data distribution to a new distribution. Therefore, we use the adversarial AE (AAE)\cite{makhzani2015adversarial} structure to transform the measured data latent features to the corresponding simulated data latent features. We also incorporate the latent space regularization scheme of variational AE (VAE)\cite{kingma2013auto} to improve the robustness and generalisation ability of the model. Finally, the FMNet can generate cleaned-up/enhanced $\mu$-DS according to the transformed latent features. For training the network, we only used small amount of measured-simulated pairs without any additional information. This reduces the complexity of obtaining sufficient training data. Finally, We trained the network with data collected in a residential setting and tested it with data collected from the same scenario and another through-the-wall scenario. We also compared the changes in the latent space before and after applying our framework. From the qualitative and quantitative analysis, we can see the promising enhancement performance. Additionally, based on our framework, we propose a new strategy for improving $\mu$-DS classification which trains a classifier with only simulated data, and the new measured samples will be classified after being cleaned up with the FMNet.

The rest content of our paper is organized as follows: Section \ref{sec: ProblemSetting} describes current issues, then we introduced some possible solutions and their limitations. Next, Section \ref{sec: ProposedMethod} presents the proposed FMNet and the three-phase training scheme. Then in Section \ref{sec: EnhancementExperiments}, we introduced relevant systems, the experimental setting and neural network structures, etc. After that, we qualitatively and quantitatively evaluated the FMNet from various aspects and compared it with other networks in Section \ref{sec: EnhancementEvaluation}. Next in Section \ref{sec: Classification}, we introduced and evaluated the proposed simulated data-based classifier training scheme to further demonstrate the ability of the FMNet. Finally, we concluded the paper in Section \ref{sec: Conclusion}.

\section{Problem Setting}
\label{sec: ProblemSetting}
In a radar sensing task, we have a set of measured samples $M=\{m^{1},m^{2},...,m^{i}\}$ and a matched set of simulated samples $S=\{s^{1},s^{2},...,s^{i}\}$. The clean version for each $m^{i}$ is $s^{i}$ which only contains the target's motion information. In this case, it appears that the enhancement task can be processed as the supervised learning problem where we minimize the distance between the label $s^{i}$ and predicted output $m^{i'}$ through convolutional neural networks (CNNs). For this objective, neural networks require a significant quantity of training data to prevent various environmental clutter impacting model generalisation ability. However, acquisition of radar data is time-consuming and laborious. Therefore, instead of mapping $m^{i}$ to $s^{i}$, we consider to find the similarity of their latent features. 

Let $E_{\theta_{1}}$ represent the Encoder model parametrized by $\theta_{1}$ and $D_{\theta_{2}}$ represent the Decoder model parametrized by $\theta_{2}$. When training these models with $n$ simulated samples, we need to minimize the reconstruction loss:
\begin{equation}
\label{AE_Lrec}
    L_{rec}(\theta_{1},\theta_{2}) = \frac{1}{n}\sum_{i=1}^{i=n}(D_{\theta_{2}}(E_{\theta_{1}}(s^{i}))-s^{i})^2
\end{equation}
so that $D_{\theta_{2}}$ can output reconstructed data $s^{i}_{rec}$ based on the latent features $z_{s}^{i}$ extracted from $E_{\theta_{1}}$, and $s^{i}_{rec}$ will be almost identical to $s^{i}$. However, if we now feed $m^{i}$ into the models, the output is likely to be nonsensical. This is because that even for the same activity, noise in the $m^{i}$ will cause its latent features $z_{m}^{i}$ to be different from $z_{s}^{i}$. Eventually, $D_{\theta_{2}}$ may fail to derive the desired output from $z_{m}^{i}$ as the model has not been trained with this kind of data. But since $m^{i}$ and $s^{i}$ contain the same motion information, we can possibly minimize the difference between $z_{m}^{i}$ and $z_{s}^{i}$. Formally we would like to map $p(z|m^{i})$ to $p(z|s^{i})$, where $p(z|m^{i})$ is the distribution of latent features of the measured sample and $p(z|s^{i})$ is the distribution of latent features of the simulated sample. As a result, $D_{\theta_{2}}$ can draw a good quality $\mu$-DS just like $s^{i}$ from $p(z|m^{i})$ after the mapping. 

\subsection{Generative Adversarial Networks}
The Generative Adversarial Networks (GANs)\cite{goodfellow2014generative} can map one distribution to another with its adversarial structure. It consists of two sub-networks: a Generator $G$ and a Discriminator $D$. The $G$ tries to map samples $x$ from the prior distribution $p(x)$ to a desired distribution $q(x)$. Simultaneously, the $D$ exams whether the outputs of $G$ follow $q(x)$. They play a minmax two-player game and achieve the goal through the competition. The loss function of GANs can be expressed as:
\begin{dmath}
\label{basic_gan_loss}
\min_{\text{G}}\max_{\text{D}}L=E_{x\sim q(x)}[logD(x)]\\+E_{z\sim pr(z)}[log(1-D(G(z)))] 
\end{dmath}
where $z$ is the random noise sampled from distribution $p_{r}$.

We are specially interested in how the adversarial structure can help map $p(z|m^{i})$ to $p(z|s^{i})$ in the AE network, where we hope to discover the similarity between $z_{m}^{i}$ and $z_{s}^{i}$ by minimizing the adversarial loss.

\subsection{Adversarial AutoEncoder Structure}
\label{B:AAE}
\begin{figure}
\centering
\includegraphics[width=\linewidth]{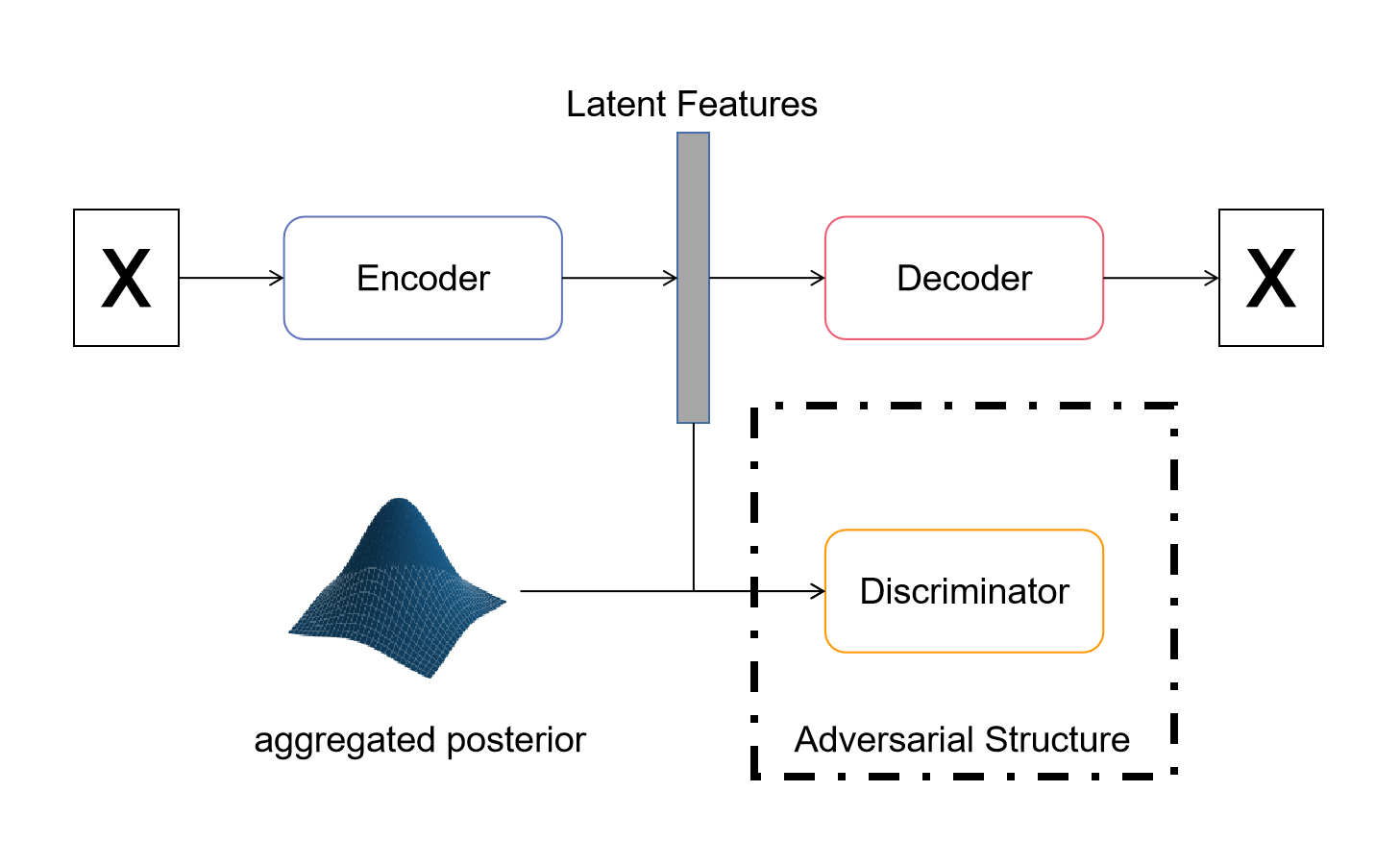}
\caption{The network structure of an AAE}
\label{fig: AAE}
\end{figure}
An effective technique that combines the adversarial structure with the AE is the AAE proposed by Makhzani et al.~\cite{makhzani2015adversarial}, and now has been applied in denoising\cite{creswell2018denoising}, data clustering\cite{ge2019dual} and many other applications. The AAE is a generative network, which aims to regularize the latent space by matching the aggregated posterior, normally Gaussian distribution, to the arbitrary prior distribution of latent features so that the decoder can transform any feature vectors of the latent space to a meaningful content. An adversarial network attached on the top of latent features is the key to achieve that, as shown in Fig. \ref{fig: AAE}. 

The AAE's training consists of two phases: reconstruction phase and regularization phase. The reconstruction phase acts just like the conventional AE aiming to minimize the distance between the reconstructed data and the original data. On the other hand, the encoder $E$ and an adversarial network construct a GAN-like structure during the regularization phase. Their roles are comparable to $G$ and $D$ of GANs, respectively. We first produce fake samples (i.e. latent feature vectors) from $E$ and then get real samples from a Gaussian distribution. Finally, the arbitrary prior distribution of latent features can be properly matched to the Gaussian distribution by minimising the adversarial loss.

This inspires us to consider replacing the Gaussian posterior with $p(z|s^{i})$ so that we can induce the prior $p(z|s^{i})$ to match it. However, $p(z|s^{i})$ is an unknown distribution, resulting in a risk of losing a well-regularized latent space which negatively impacts the model's generalisation ability. 

\subsection{Variational AutoEncoder}
\label{subsec: VAE}
The VAE is a popular generative network. The regularization phase of the VAE proposed by Kingma and Welling~\cite{kingma2013auto} is similar to AAEs, where the aggregated posterior is normally assumed to be Gaussian $q(z|x)$ when the input is $x$. The VAE minimizes the difference between $q(z|x)$ and an unknown prior $p(z|x)$ measuring by the Kullback-Leibler divergence (KLD). The objective function of the network is composed of the reconstruction $L_{rec}$ and regularization $L_{kl}$ terms, which can be simply expressed as:
\begin{dmath}
\label{VAE_loss}
L_{vae}=-E_{q(z|x)}[log(p(z|x))]+\mathbb{KL}(q(z|x)||p(z|x))
\end{dmath}
where $\mathbb{KL}$ calculates the KLD between two distributions.

In practice, the VAE describes $i^{th}$ observations of latent features as a function of mean $\mu_{i}$ and standard derivation $\sigma_{i}$ of Gaussian $N(\mu_{i},\sigma_{i})$ and applies backpropogation on $\mu_{i}$ and $\sigma_{i}$ by using reparameterization trick:
\begin{equation}
\label{repara}
z=\mu+\epsilon\sigma, \text{where }\epsilon~N(0, 1)
\end{equation}
This restricts $p(z|x)$ must be described in functional form. As a result, we cannot map $p(z|m^{i})$ to the arbitrary distribution $p(z|s^{i})$ with the VAE. However, the regularization method can be incorporated into the AAE structure and this may solve the difficulty we discussed at the end of Section \ref{B:AAE}.

\section{Proposed Enhancement Network}
\label{sec: ProposedMethod}
\begin{figure*}
\centering
\includegraphics[width=\linewidth]{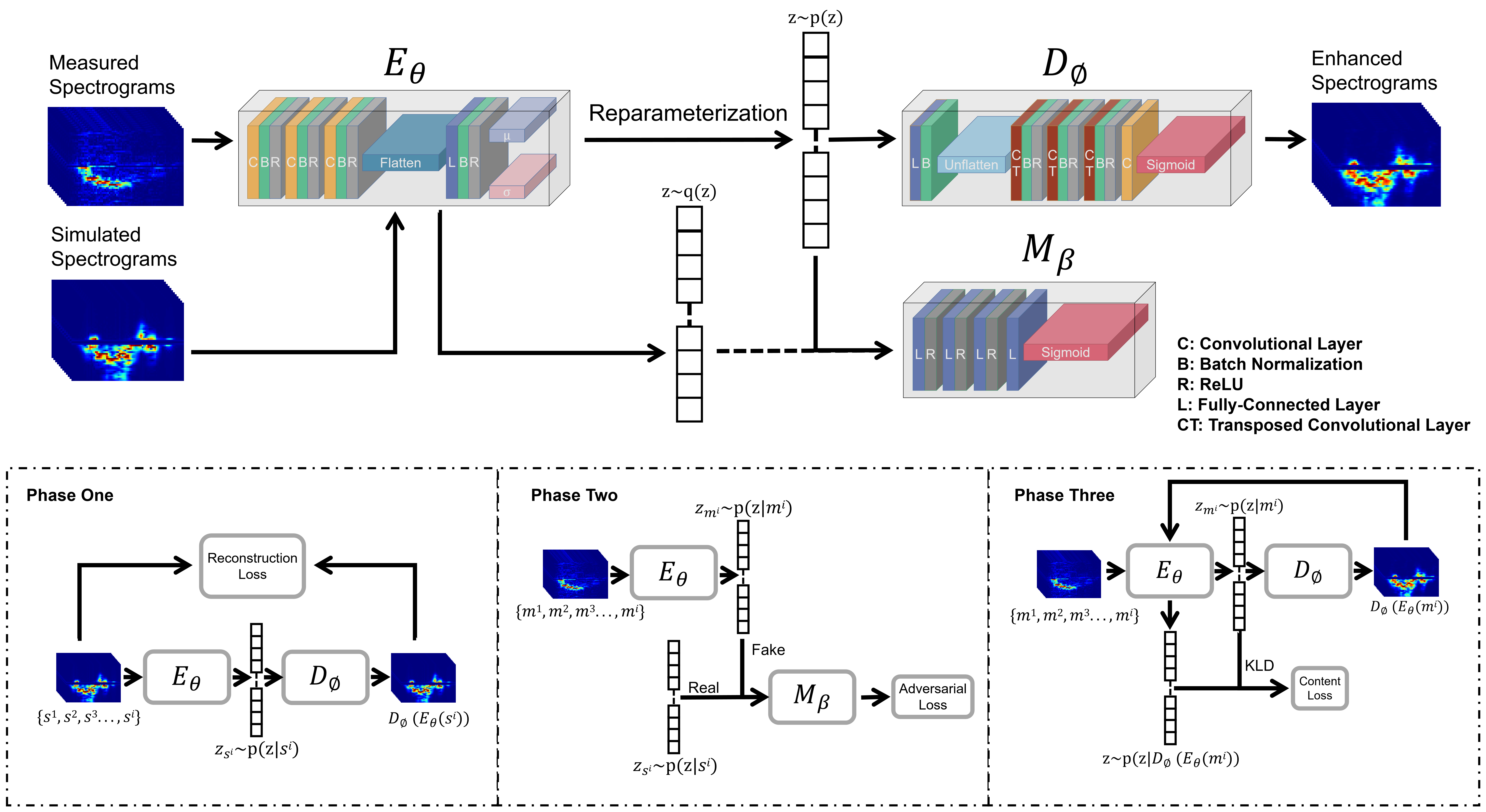}
\caption{The proposed network structure and different training phases}
\label{fig: proposed network}
\end{figure*}
When using the AE structure to enhance $\mu$-DS, it acts like a generative network where the decoder generates new samples based on latent features. As a result, the network needs a well-regularized latent space as discussed in Section \ref{B:AAE} and \ref{subsec: VAE}. Meanwhile, to maximize the motion information similarity of measured data and matched simulated data through mapping $p(z|m^{i})$ to $p(z|s^{i})$, the adversarial structure can be very helpful. After taking into account these considerations, our proposed FMNet attaches an adversarial network on a single VAE structure to achieve $\mu$-DS enhancement task, as shown in Fig. \ref{fig: proposed network}.

The FMNet structure can eventually provide us:
\begin{itemize}
    \item a strong encoder $E_{\theta}$ parametrized by $\theta$ capable of extracting equivalent latent features from measured and matched simulated spectrograms for the same activities.
    \item a discriminator $M_{\beta}$ parametrized by $\beta$ helping us achieve feature mapping.
    \item a well-regularized latent space ensuring that the decoder can generate meaningful content from any latent regions.
    \item a robust decoder $D_{\phi}$ parametrized by $\phi$ capable of producing a simulated-like enhanced spectrogram based on latent features extracted from a measured $\mu$-DS.
\end{itemize}
Furthermore, the enhanced output should retain the most of the motion information of input data, rather than merely a random sample from the same activity distribution. During the training, we design a three-phases scheme to gradually achieve above goals.

\subsection{Phase One: Reconstruction and Regularization}
\label{subsec: phase one}
This phase ignores the existence of $M_{\beta}$ and aims to train both the $E_{\theta}$ and the $D_{\phi}$ through minimizing the reconstruction loss. Specifically, we use $S$ as the training set, because we only need $D_{\phi}$ to generate data from the latent space of simulated samples. Furthermore, the VAE regularization scheme will be applied at the end of encoding stage. The objective of this phase is to minimize the following loss function:
\begin{dmath}
\label{phase1_loss}
L_{1}=\frac{1}{n}\sum_{i=1}^{i=n}(||D_{\phi}(E_{\theta}(s^{i}))-s^{i}||^2 \\+ \mathbb{KL}[N(\boldsymbol{\mu},\boldsymbol{\sigma}),N(\boldsymbol{0}, \boldsymbol{I})])
\end{dmath}
where $N(\boldsymbol{\mu},\boldsymbol{\sigma})$ is normal distributions defined by mean $\boldsymbol{\mu}$ and standard derivation $\boldsymbol{\sigma}$ diagonal matrices of latent features. The size of matrices is decided by the number of latent features. $N(\boldsymbol{0}, \boldsymbol{I})$ is standard normal distributions and $\boldsymbol{I}$ is the identity matrix with the same size as $\boldsymbol{\mu}$ and $\boldsymbol{\sigma}$.

The first term in the loss function is the reconstruction loss which can be calculated by the mean square error (MSE). For the second term, it calculates the KLD between a normal distribution of one latent feature and a standard normal distribution for the regularization purpose.

\subsection{Phase Two: Feature Mapping}
In this phase, we freeze $D_{\phi}$ and train $E_{\theta}$ and $M_{\beta}$. These two networks form a GAN structure, and for our feature mapping purpose, we can specifically describe the objective function of the GAN as:
\begin{dmath}
\label{proposed gan loss}
\min_{E_{\theta}}\max_{M_{\beta}}L=E_{z\sim p(z|s^{i})}[log M_{\beta}(z)]\\+E_{z\sim p(z|m^{i})}[log (1-M_{\beta}(E_{\theta}(z)))]   
\end{dmath}
The GAN training requires \"real\" and \"fake\" pairs. The real sample in this case is the latent feature vector of simulated $\mu$-DS, whereas the fake sample is the latent feature vector of measured $\mu$-DS. Specifically, after obtaining the $E_{\theta}$ from the previous phase, we can respectively extract the real and the fake samples through $z_{s}^{i}=E_{\theta}(s^{i})$ and $z_{m}^{i}=E_{\theta}(m^{i})$ for the $i^{th}$ simulated and measured $\mu$-DS pair. The training set will then be a collection of many $(z_{s}^{i}, z_{m}^{i})$ pairs that have been labelled with $(Real, Fake)$. Next, $M_{\beta}$ receives these samples and acts like a discriminator. 

Overall, the training process is same as in \cite{goodfellow2014generative}. Notably, we should use matched simulated and measured spectrograms as pairs to ensure that the motion information expressed by them is consistent.

\subsection{Phase Three: Content Consistency Check}
The content consistency check is required for the final step because we need to confirm that the enhanced output and the original input express the same motion information. However, due to various noise, the traditional evaluation methods, such as MSE and the Structural Similarity Index Measure (SSIM), cannot obtain the desired comparison result. Therefore, we feed the output back into $E_{\theta}$ and check their content consistency by measuring the KLD of the distributions of their latent features.

Assume a measured data $m^{i}$ is passed through $E_{\theta}$ and $D_{\phi}$, we will get the latent features: $z_{m^{i}}=E_{\theta}(m^{i})$ and the enhanced data: $m^{i'}=D_{\phi}(z_{m^{i}})$. Then feeding $m^{i'}$ back to $E_{\theta}$, we can get its latent features: $z_{m^{i'}}=E_{\theta}(m^{i'})$. As mentioned before, each observations of $z_{m^{i}}$ and $z_{m^{i'}}$ are sampled from a Gaussian. Let $N(\mu_{j},\sigma_{j})$ and $N(\mu_{j}^{'},\sigma_{j}^{'})$ denote the distributions of $j^{th}$ observation of $z_{m^{i}}$ and $z_{m^{i'}}$, respectively. The objective function of this phase can be expressed as:
\begin{dmath}
\label{content check}
L_{content}=\frac{1}{n}\sum_{j=1}^{j=n}\mathbb{KL}[N(\mu_{j},\sigma_{j}),N(\mu_{j}^{'},\sigma_{j}^{'})]  
\end{dmath}
where $n$ is the number of observations in the latent features. Finally, by optimizing Equation \ref{content check}, we can ensure that the enhanced result contains the close motion information as the measured input because their latent features extracted by $E_{\theta}$ are similar to each other in terms of the probability distribution.

\section{Enhancement Experiments}
\label{sec: EnhancementExperiments}
\begin{figure}
\centering
\includegraphics[width=\linewidth]{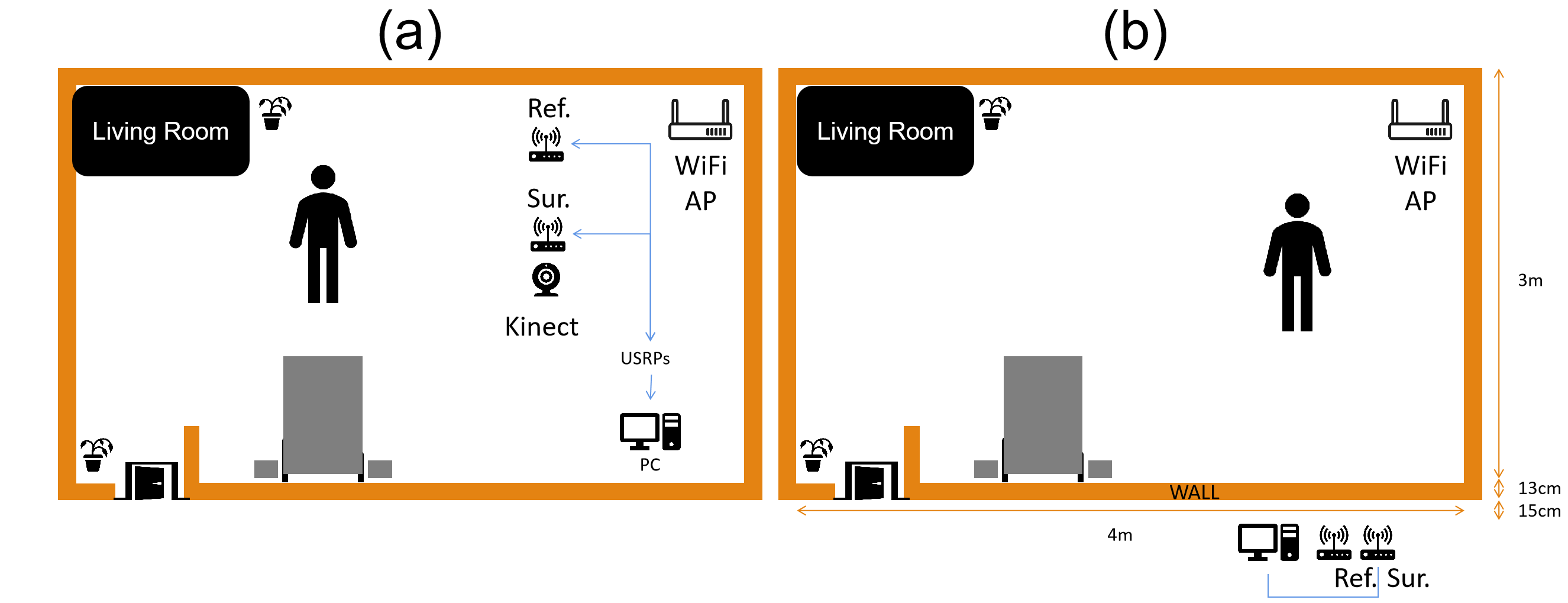}
\caption{(a). The schematics of the Line of Sight (LOS) experimental scene and (b). the TTW experimental scene: two scenes refer to the same room with size of $4m$x$3m$, the wall thickness is $13cm$, the stand off distance between wall and antennas in TTW scene is $15cm$}
\label{fig: exp_setup}
\end{figure}
Experiments were designed in various scenarios to verify the performance of the FMNet. During the experiments, the Passive WiFi Radar (PWR) system is utilised to acquire the measured $\mu$-DS, while a Kinect sensor simultaneously records motion capture (Mocap) data, which is subsequently utilised to generate the simulation data. Figure \ref{fig: exp_setup} (a) illustrates how the two systems cooperates together.

In this section, we will briefly introduce the PWR system and our simulation software, SimHumalator, as well as go through the dataset and proposed networks in further depth.

\subsection{Passive WiFi Radar System}
\label{subsec: PWR}
As shown in Figure \ref{fig: exp_setup},  the PWR system in experiments consists of two Yagi antennas- one acting as the reference channel and another one acting as a surveillance channel. The reference channel receives direct transmissions, and the surveillance channel gathers signals reflected off the targets. Then, signals from two channels are processed based on the cross-correlation to generate the measured $\mu$-DS. Furthermore, two NI USRP-2921 software-defined radios are connected to the two channels for the channel synchronization and real-time signal acquisition purposes. See more detains about the PWR system and its signal processing method in \cite{li2020passive}. 

\subsection{SimHumalator}
\label{subsec: Simhumalator}
\begin{figure}
\centering
\includegraphics[width=\linewidth]{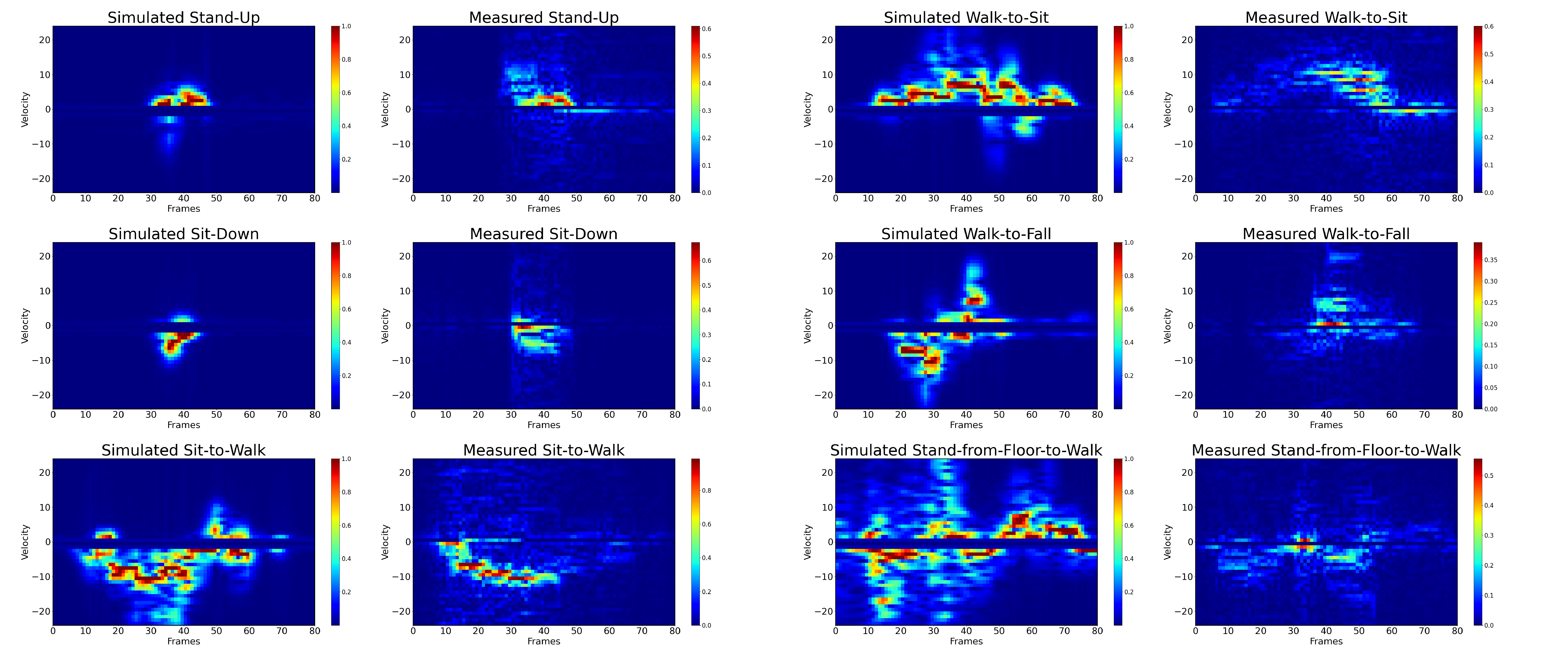}
\caption{Examples of the measured and simulated $\mu$-DS pairs}
\label{fig: sim egs}
\end{figure}
SimHumalator is an opensource simulation framework for generating human $\mu$-DS in the PWR sensing scenarios and can be downloaded for free from \url{https://https://uwsl.co.uk/}. It simulates the IEEE 802.11g standard WiFi transmissions using MATLAB's WLAN toolbox, and human animations based on the Mocap data. In the experiments, a Kinect sensor that was synchronised with the PWR system was positioned in the same location as the surveillance channel, ensuring that the measured and simulated spectrograms expressed the identical motion information. Figure \ref{fig: sim egs} shows the consistency of measured and simulated $\mu$-DS. See more details on SimHumalator in \cite{vishwakarma2021simhumalator}.

\subsection{Dataset Explanation}
\label{subsec: dataset}
We gathered data from three participants, each performing six different activities- sit-down, stand-up, sit-to-walk, walk-to-sit, walk to fall-down, and stand-from-floor to walk. Each activities is completed within 5-10 seconds and repeated 15-20 times, resulting in average 60 $\mu$-DS and Mocap data for each activity. Following this, the Mocap data was used to produce simulated data with SimHumalator that corresponded to the measured spectrogram. Finally, we obtained 304 measured $\mu$-DS and 304 matched simulated $\mu$-DS.

Data were also collected for through-the-wall (TTW) scenario of activities- sit-down, stand-up, sit-to-walk and walk-to-sit, to further demonstrate the performance of the proposed framework under different sensing environments.

\subsection{Network and Training Setting}
\label{subsec: network details}
\begin{table*}[]
\resizebox{\textwidth}{!}{%
\begin{tabular}{|c|c|c|c|c|}
\multicolumn{5}{l}{Encoder Structure} \\ \hline
Name of Layer & Type of Layer & Parameter Setting & Tensor Shape & Previous Layer \\ \hline
EC1 & Conv2D & filters 32, kernel size 5x5,  stride 2x2, padding 0, BatchNormalization, ReLU & (24, 40, 32) &  \\ \hline
EC2 & Conv2D & filters 64, kernel size 5x5,  stride 2x2, padding 0, BatchNormalization, ReLU & (12, 20, 64) & EC1 \\ \hline
EC3 & Conv2D & filters 128, kernel size 5x5,  stride 2x2, padding 0, BatchNormalization, ReLU & (6, 10, 128) & EC2 \\ \hline
EF & Flatten &  & (7680,) & EC3 \\ \hline
EL1 & Linear & in features 7680, out features 1024, BatchNormalization, ReLU & (1024,) & EF \\ \hline
mu & Linear & in features 1024, out features 2048 & (2048,) & EL1 \\ \hline
var & Linear & in features 1024, out features 2048 & (2048,) & EL1 \\ \hline
\multicolumn{5}{|c|}{Reparameterization} \\ \hline
\multicolumn{5}{l}{Decoder Structure} \\ \hline
Name of Layer & Type of Layer & Parameter Setting & Tensor Shape & Previous Layer \\ \hline
DL1 & Linear & in features 2048, out features 7680, BatchNormalization & (7680,) &  \\ \hline
DUF & Unflatten &  & (6, 10, 128) & DL1 \\ \hline
DCT1 & TransposedConv2D & filters 64, kernel size 5x5,  stride 2x2, padding 0, BatchNormalization, ReLU & (12, 20, 128) & DUF \\ \hline
DCT2 & TransposedConv2D & filters 32, kernel size 5x5,  stride 2x2, padding 0, BatchNormalization, ReLU & (24, 40, 128) & DCT1 \\ \hline
DCT3 & TransposedConv2D & filters 16, kernel size 5x5,  stride 2x2, padding 0, BatchNormalization, ReLU & (48, 80, 128) & DCT2 \\ \hline
DC1 & Conv2D & filters 1, kernel size 5x5,  stride 1x1, padding 0, Sigmoid & (48,80,1) & DCT3 \\ \hline
\multicolumn{5}{l}{Discriminator Structure} \\ \hline
Name of Layer & Type of Layer & Parameter Setting & Tensor Shape & Previous Layer \\ \hline
DSL1 & Linear & in features 2048, out features 1000, ReLU & (1000,) &  \\ \hline
DSL2 & Linear & in features 1000, out features 500, ReLU & (500,) & DSL1 \\ \hline
DSL3 & Linear & in features 500, out features 215, ReLU & (215,) & DSL2 \\ \hline
DSL4 & Linear & in features 215, out features 1, ReLU & (1,) & DSL3 \\ \hline
\end{tabular}%
}
\caption{Neural Network Structures of Encoder, Decoder and Discriminator: The input $\mu$-DS shape is $(48,80,1)$}
\label{tab:networks sturcture}
\end{table*}

The FMNet is implemented using Pytorch library\cite{paszke2017automatic} on NVidia 1060 GPU card. It consists of three parts- an encoder, a decoder and a discriminator. Their structural details are presented in Table \ref{tab:networks sturcture}. 

\section{Enhancement Performance Evaluations}
\label{sec: EnhancementEvaluation}
\begin{figure*}
\centering
\includegraphics[width=\linewidth]{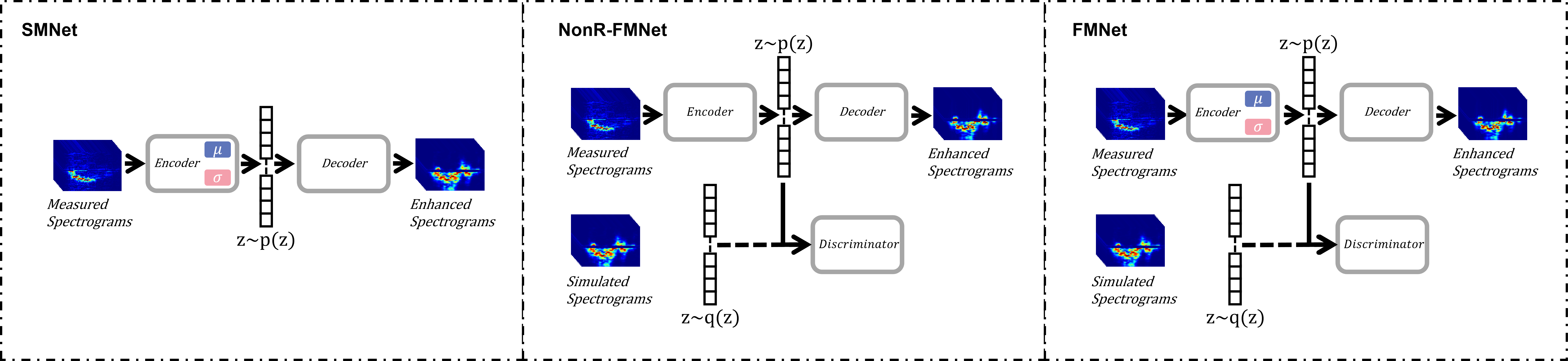}
\caption{The comparison of network structures of SMNet, NonR-FMNet and FMNet}
\label{fig: three networks}
\end{figure*}
To demonstrate the performance of our enhancement method, we compared the FMNet with other two types of networks which have different structures and training schemes. The first one is based on the VAE network\cite{kingma2013auto} and trains the model with the conventional supervised learning method i.e. the simulated samples are labels of the measured samples. We named it as the Spectrogram Mapping network (SMNet). The second network has the similar strucuter to the AAE\cite{makhzani2015adversarial}. We named it as the Nonregularized Feature Mapping network (NonR-FMNet), which uses a similar training technique to FMNet but does not regularize latent space. The structural comparison of three networks has been shown in Fig. \ref{fig: three networks}.
\subsection{Visualizing Enhancement Performance}
\begin{figure*}
\centering
\includegraphics[width=\linewidth]{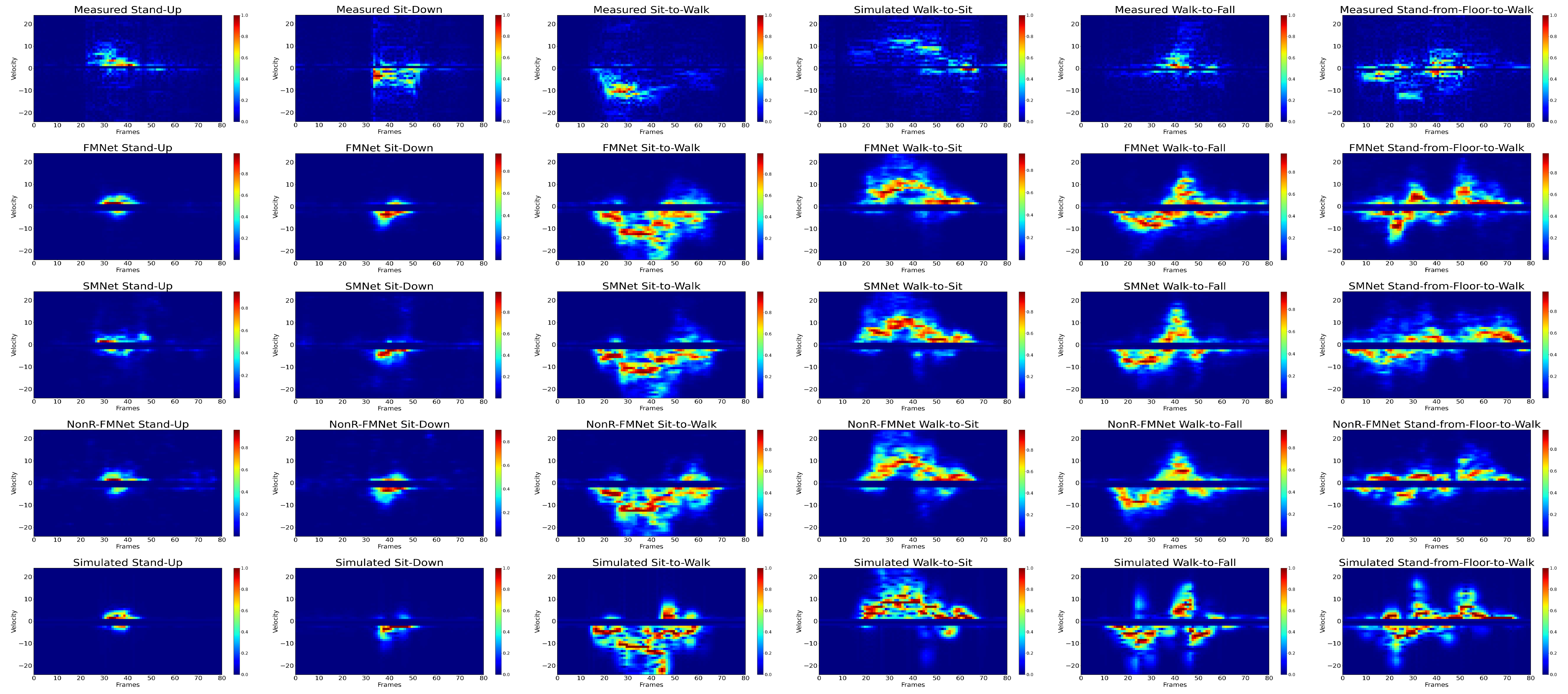}
\caption{The first row presents the measured spectrograms of six activities; the last row presents the matched simulated spectrograms; the rest rows present enhanced spectrograms corresponding to the measured data}
\label{fig: result1}
\end{figure*}
We can observe in Fig. \ref{fig: result1} that all three networks can produce a clean $\mu$-DS from the original measured data. In terms of the qualitative analysis, the results of FMNet have distinct patterns and features, while the stand-up, sit-down, stand-from-floor-to-walk results of the SMNet and NonR-FMNet have blurred areas in the background and their similarity to the simulated spectrogram is not high. This is because that the direct mapping method of the SMNet is quite challenging and with the small amount of training data, it only has the limited performance. For the NonR-FMNet, it has the discontinuity in the latent space. When a latent observation falls into discontinuous areas, the decoder cannot properly interpret it, leading to producing some nonsensical blurs. Furthermore, when the SMNet and FMNet performance are compared, we can see that the FMNet results are closer to the corresponding simulated data (like the stand-from-floor-to-walk case). In contrast, while the SMNet outputs are of the same class as the corresponding simulated data, there are many differences that imply their motion information may not be the same. Actually, the SMNet training strategy merely takes into account the global relationship between the measured and generated spectrograms. When there aren't enough training cases, the network may struggle to account for new changes. Intuitively, we can see that the FMNet outperforms the other two networks in terms of clarity and similarity. On the other hand, although the results from the other networks are slightly worse, they are also acceptable. So for the clearer comparison, we used TTW dataset to further test the generalisation ability of three networks. 

\subsection{generalisation Ability Evaluation}
\begin{figure*}
\centering
\includegraphics[width=\linewidth]{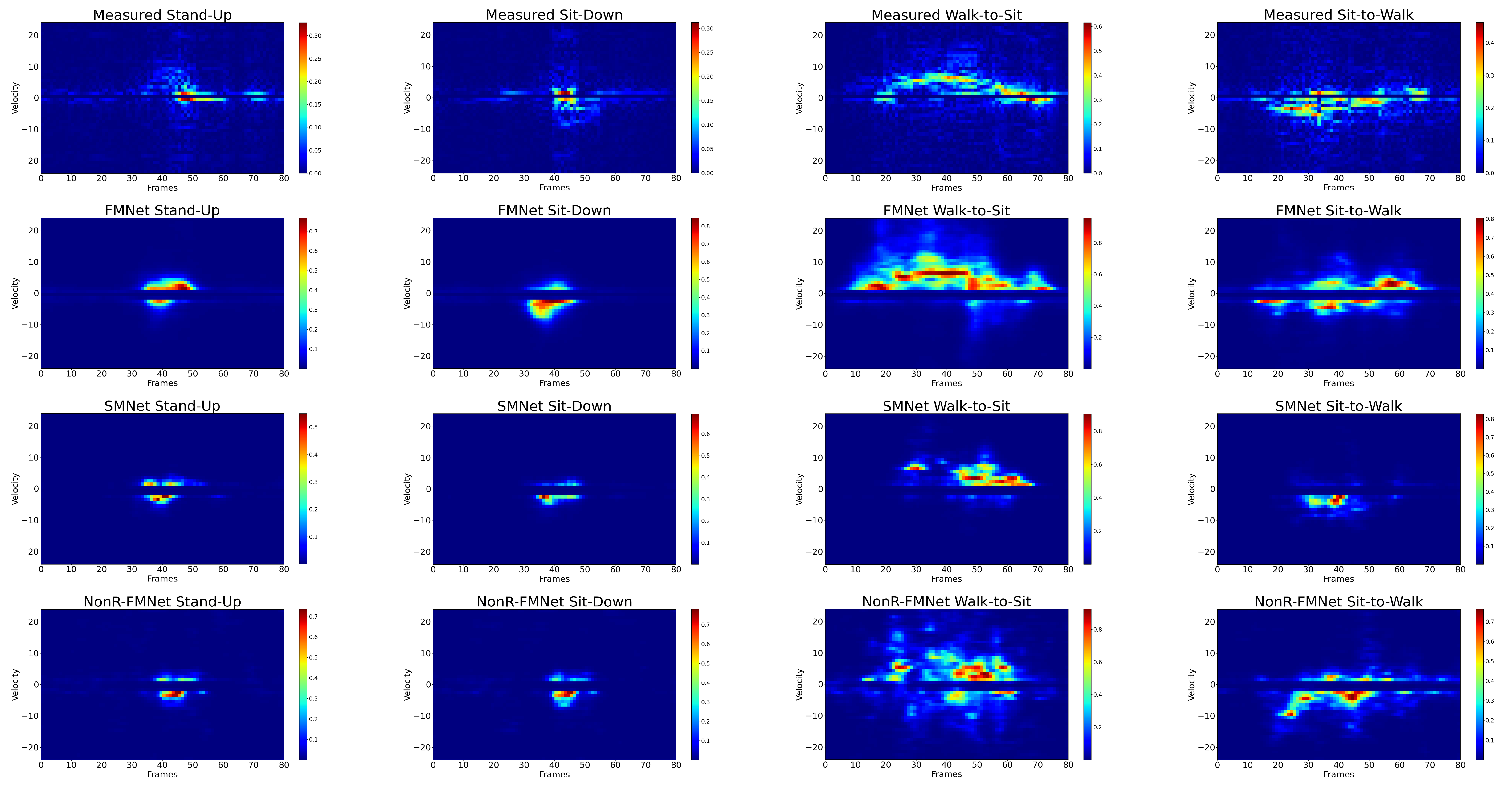}
\caption{The first row presents the TTW measured spectrograms of four activities; the rest rows present the enhanced results corresponding to the measured data}
\label{fig: TTW results}
\end{figure*}
The TTW experimental scenario has been shown in Fig. \ref{fig: exp_setup} (b), and the enhancement results are presented in Fig. \ref{fig: TTW results}. We can observe that only FMNet can use the same network to successfully produce enhanced spectrograms of the TTW measured data. For the NonR-FMNet and SMNet, except the sit-down case, they either generated meaningless content (like sit-to-walk and walk-to-sit cases) or generated results that do not match the measured input (like stand-up cases). Furthermore, even though the walk-to-sit spectrogram in the TTW scenario differs considerably from the one in the previous dataset, the FMNet can still enhance the spectrogram based on the provided motion pattern to the greatest extent, rather than generating meaningless content like other approaches did. By comparing motion patterns between enhanced outputs and the measured data, the FMNet shows the robust generalisation ability.

\subsection{Analysis of Latent Space}
\label{latent space analysis}
\begin{figure*}
\centering
\includegraphics[width=\linewidth]{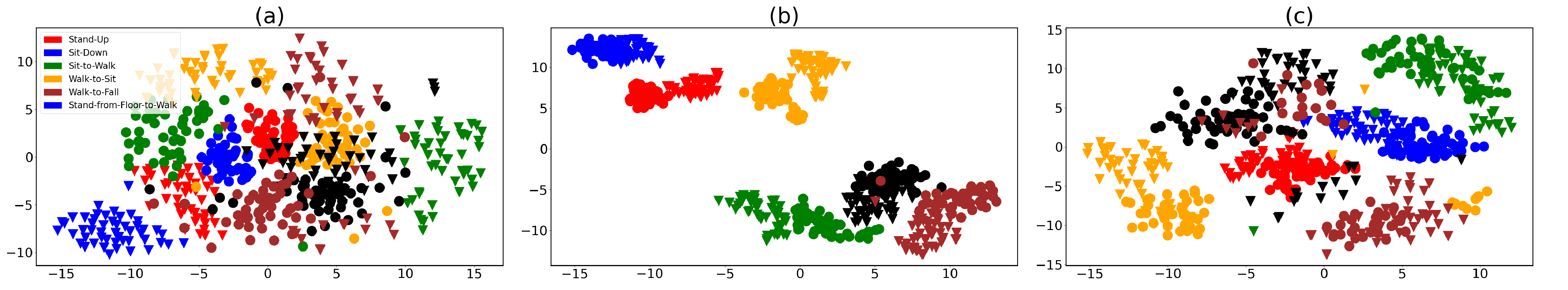}
\caption{t-SNE projections of latent features of (a) SMNet, (b) NonR-FMNet and (c) FMNet cases: the circle markers represent latent features of the measured samples, and the triangle markers represent latent features of the simulated samples}
\label{fig: tsne plots}
\end{figure*}
To study changes happened in the latent space, we fed some measured and simulated spectrogram samples into the encoders of the three networks and then inspected the latent space with t-SNE\cite{van2008visualizing} plots shown in Fig. \ref{fig: tsne plots}. We can observe different properties of these latent spaces. The distributions of both measured and simulated data are self-clustered in all cases. Furthermore, clusters in the NonR-FMNet and FMNet cases from the same class are near to one another and separable from the other classes, which indicates latent features carry information for the classification. Also this demonstrates that our feature mapping approach can successfully discover the latent similarity between the measured and simulated data and map features together. However, clusters in the NonR-FMNet are more concentrated, causing the presence of empty spaces between different classes, while feature points in the FMNet are evenly distributed throughout the latent space. As previously mentioned, the latter characteristic is more beneficial to our generative network. On the other hand, we can also observe different classes are clustered in the latent space of the SMNet, but they are not separable. Meanwhile, the measured and simulated data sharing the same motion information did not grouped together. This suggests that the SMNet did not discover the similarity between the matched data. From this difference, we can see the superiority of the proposed feature mapping structure.

\begin{table}[]
\resizebox{\linewidth}{!}{%
\begin{tabular}{cc|c}
\cline{2-3}
 & FMNet & SMNet \\ \hline
\multicolumn{1}{c|}{Stand-Up} & 5.41e-02 & 1.72e+09 \\ \hline
\multicolumn{1}{c|}{Sit-Down} & 4.66e-02 & 2.43e+09 \\ \hline
\multicolumn{1}{c|}{Sit-to-Walk} & 2.00e-01 & 1.40e+11 \\ \hline
\multicolumn{1}{c|}{Walk-to-Sit} & 1.63e-01 & 1.95e+10 \\ \hline
\multicolumn{1}{c|}{Walk-to-Fall} & 3.33e-01 & 9.92e+09 \\ \hline
\multicolumn{1}{c|}{Stand-from-Floor-to-Walk} & 2.38e-01 & 9.72e+11 \\ \hline
\end{tabular}%
}
\caption{KLD comparison between FMNet and SMNet: each value is the average of $2048$ KLD values from the measured data latent features and the simulated data latent features }
\label{tab:KLD}
\end{table}
To quantitatively study their latent spaces, we also used Equation \ref{content check} to measure the multivariate KLD of latent feature distributions of measured and simulated data pairs for the FMNet and SMNet cases, as shown in Table \ref{tab:KLD}. The results are consistent with the above analysis. With the feature mapping structure, the KLD of the FMNet cases is much smaller than it of the SMNet case. Although the training mechanisms of these networks are different, this still can demonstrates the changes brought by the feature mapping technique and it works very well in the real applications. 

\subsection{Structural Similarity and Content Difference}

\begin{table}[]
\resizebox{\linewidth}{!}{%
\begin{tabular}{ccc|c|c|c}
\cline{3-6}
 &  & Mear vs Sim & FMNet vs Sim & SMNet vs Sim & NonR-FMNet vs Sim \\ \hline
\multicolumn{1}{c|}{\multirow{2}{*}{Stand-Up}} & \multicolumn{1}{c|}{Pixel Loss} & 0.0045 & \textbf{0.0006} & 0.0010 & 0.0018 \\ \cline{2-6} 
\multicolumn{1}{c|}{} & \multicolumn{1}{c|}{SSIM} & 0.7401 & \textbf{0.9549} & 0.9519 & 0.9136 \\ \hline
\multicolumn{1}{c|}{\multirow{2}{*}{Sit-Down}} & \multicolumn{1}{c|}{Pixel Loss} & 0.0050 & \textbf{0.0004} & 0.0009 & 0.0016 \\ \cline{2-6} 
\multicolumn{1}{c|}{} & \multicolumn{1}{c|}{SSIM} & 0.7998 & \textbf{0.9625} & 0.9596 & 0.9258 \\ \hline
\multicolumn{1}{c|}{\multirow{2}{*}{Sit-to-Walk}} & \multicolumn{1}{c|}{Pixel Loss} & 0.0341 & \textbf{0.0028} & 0.0055 & 0.0117 \\ \cline{2-6} 
\multicolumn{1}{c|}{} & \multicolumn{1}{c|}{SSIM} & 0.4541 & \textbf{0.8942} & 0.8670 & 0.7460 \\ \hline
\multicolumn{1}{c|}{\multirow{2}{*}{Walk-to-Sit}} & \multicolumn{1}{c|}{Pixel Loss} & 0.0328 & \textbf{0.0049} & 0.0058 & 0.0110 \\ \cline{2-6} 
\multicolumn{1}{c|}{} & \multicolumn{1}{c|}{SSIM} & 0.4532 & 0.8435 & \textbf{0.8554} & 0.7482 \\ \hline
\multicolumn{1}{c|}{\multirow{2}{*}{Walk-to-Fall}} & \multicolumn{1}{c|}{Pixel Loss} & 0.0325 & \textbf{0.0052} & 0.0087 & 0.0159 \\ \cline{2-6} 
\multicolumn{1}{c|}{} & \multicolumn{1}{c|}{SSIM} & 0.5243 & \textbf{0.8698} & 0.7712 & 0.6498 \\ \hline
\multicolumn{1}{c|}{\multirow{2}{*}{Stand-from-Floor-to-Walk}} & \multicolumn{1}{c|}{Pixel Loss} & 0.0458 & 0.0094 & \textbf{0.0090} & 0.0173 \\ \cline{2-6} 
\multicolumn{1}{c|}{} & \multicolumn{1}{c|}{SSIM} & 0.4213 & \textbf{0.8357} & 0.8006 & 0.6542 \\ \hline
\end{tabular}%
}
\caption{Pixel loss and SSIM before and after using different networks: the Mear vs Sim column is the baseline which compares pixel loss and SSIM between the original measured data (Mear) and the simulated data (Mear); the lowest pixel loss and highest SSIM have been highlighted}
\label{tab:pl_ssim}
\end{table}

\begin{table}[]
\resizebox{\linewidth}{!}{%
\begin{tabular}{|c|c|c|c|}
\hline
 & FMNet vs Mear & SMNet vs Mear & NonR-FMNet vs Mear \\ \hline
Stand-Up & \textbf{0.8217} & 0.7707 & 0.8178 \\ \hline
Sit-Down & \textbf{0.8544} & 0.8276 & \textbf{0.8544} \\ \hline
Sit-to-Walk & \textbf{0.5031} & 0.4803 & 0.4968 \\ \hline
Walk-to-Sit & \textbf{0.6397} & 0.5307 & 0.6020 \\ \hline
\end{tabular}%
}
\caption{SSIM comparison of three networks: each value is calculated between the original TTW measured data and the corresponding enhanced spectrogram; the highest SSIM values have been highlighted}
\label{tab:ttw_ssim}
\end{table}
The previous discussions focused on whether the cleaned-up spectrograms have clear patterns, and what relationships between distributions of latent features. But what we also would like to know is if the enhanced spectrograms include the same motion information as the original measured data, and how much content is lost in the enhanced output. In this section, we use two indices to evaluate these aspects, which are Structural Similarity Index Measure (SSIM) and Pixel Loss~\cite{johnson2016perceptual}. And the pixel loss can be defined as the following.
\begin{equation}
    L_{pixel}(m,\hat{m})=\frac{||m-\hat{m}||^2}{N}
\end{equation}
where $m$ is the original measured spectrogram, $\hat{m}$ is the enhanced output and $N$ is the total number of pixels in one spectrogram.

Table \ref{tab:pl_ssim} compares the pixel loss and SSIM over three networks, where we calculated the two indices over $61$ samples and put averages in the table. For the LOS dataset, the simulated data can be regarded as the spectrogram without the environmental noise and multipath interference. We calculated $SSIM(m,s)$ and $SSIM(\hat{m},s)$, where $s$ is the simulated data. From the table,has the higher SSIM value implies that these networks can generate higher-quality enhanced spectrograms with motion patterns equivalent to their original input. Among these results, the FMNet outputs have the highest structural similarity over all cases, and compared with $SSIM(m,s)$, we can see significant improvements. Meanwhile, when comparing pixel losses, we can also observe that the FMNet has the lowest loss except walk-to-sit and stand-from-floor-to-walk cases. These observations suggest the FMNet can achieve the greatest enhancement effect while retaining more original information. For the TTW dataset, due to we did not have the simulated data, we calculated SSIM between results of three networks and the TTW measured data, as shown in Table \ref{tab:ttw_ssim}. For all activities, the FMNet results always have the highest structural similarity with the measured spectrograms, which means the FMNet can also successfully process data from different sensing scenarios and the enhanced spectrograms can express the similar content. Although the NonR-FMNet case also has the highest SSIM value for the Sit-Down activity, we can see from Figure \ref{fig: TTW results} that the result of FMNet has the stronger pattern which will be good for usages in other applications. 

\section{FMNet Application in Classification Task}
\label{sec: Classification}
The $\mu$-DS classification study is crucial for radar sensing systems in real-life applications~\cite{lei2005target, molchanov2014classification}. So far, we have shown the feasibility and robustness of the FMNet enhancement framework. In this section, we proposed a novel $\mu$-DS classification idea based on the FMNet, where a classifier can be only trained using simulated data and still achieve high classification accuracy on measured sepctrograms.

\subsection{FMNet-based Classification Scheme}
\begin{figure}
\centering
\includegraphics[width=\linewidth]{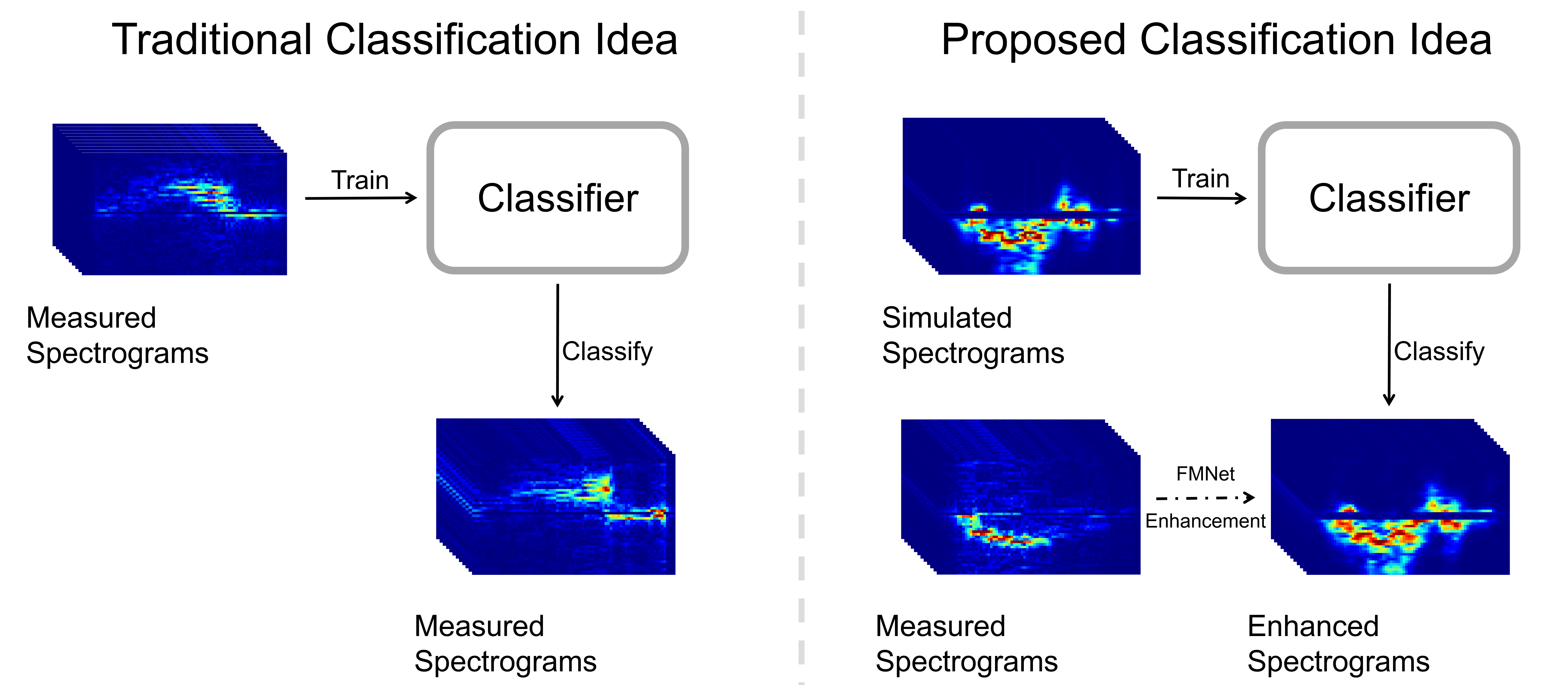}
\caption{The comparison of the traditional classification idea and the proposed classification idea: the difference is that the training set in the proposed method consists of simulated samples, and before classifying new measured samples, they are enhanced with FMNet}
\label{fig: classification}
\end{figure}
Sufficient training data is crucial for most of classification tasks. However, radar sensing scenarios are frequently plagued by a lack of training data due to the difficulty of real-world measurements. Meanwhile, unlike image data, we cannot augment data by cropping, rotation, flipping, and so on since each Doppler bin has an actual meaning related to range and speed information. Some existing $\mu$-DS augmentation methods~\cite{erol2019gan, alnujaim2019generative} used GAN-based synthetic data to augment training data. We also proposed using simulated data to the augment $\mu$-DS in \cite{tang2021augmenting}, and further improved the augmentation performance by adding GAN-generated environmental factors to the clean simulated sepctrogram in \cite{vishwakarma2021gan}. The ideas behind the aforementioned methods are similar in that they both add new samples to the training dataset in order to gain higher classification accuracy when testing on the measured samples. The benefit of using synthetic and simulated data is that we can generate a significant amount of training data easily. However, we need to ensure that the new samples approach the measured spectrograms, or that the quantity of new samples be carefully controlled so that the classifier is not confused. Actually, we can ease the issue of the lack of training data in a different way. Previously, we have demonstrated that the FMNet can produce the cleaned-up spectrograms for any measured data. Based on this fact, we can only use a large amount of simulated data to train a classifier, and then classify their enhanced spectrograms rather than the original measured spectrograms. The difference between the proposed classification method and the traditional classification method is presented in Fig. \ref{fig: classification}.

As analyzed in Section \ref{latent space analysis}, The FMNet can cluster latent features of the same classes together, regardless of whether they are from measured or simulated data. So, even if there are no measured spectrograms in the training data, the classifier can still learn unique features of each class using only simulated data. Meanwhile, the simulated data contains strong and distinct motion patterns, which is also beneficial to the training of classifier.
\subsection{Experimental Results and Evaluation}
\begin{table}[]
\resizebox{\linewidth}{!}{%
\begin{tabular}{|c|c|c|c|}
\hline
Number of Training Samples & Train OM Test OM & Train S Test EM & Train S Test OM \\ \hline
170 & 79.2\% & \textbf{87.2\%} & 58.2\% \\ \hline
194 & 81.3\% & \textbf{87.6\%} & 59.6\% \\ \hline
218 & 86.1\% & \textbf{88.9\%} & 56.7\% \\ \hline
243 & 92.1\% & \textbf{93.7\%} & 56.8\% \\ \hline
304 & - & \textbf{93.9\%} & 58.3\% \\ \hline
\end{tabular}%
}
\caption{The classification results comparison of three training schemes, the highest accuracy in each rows has been highlighted}
\label{tab:classification}
\end{table}

\begin{figure}
\centering
\includegraphics[width=\linewidth]{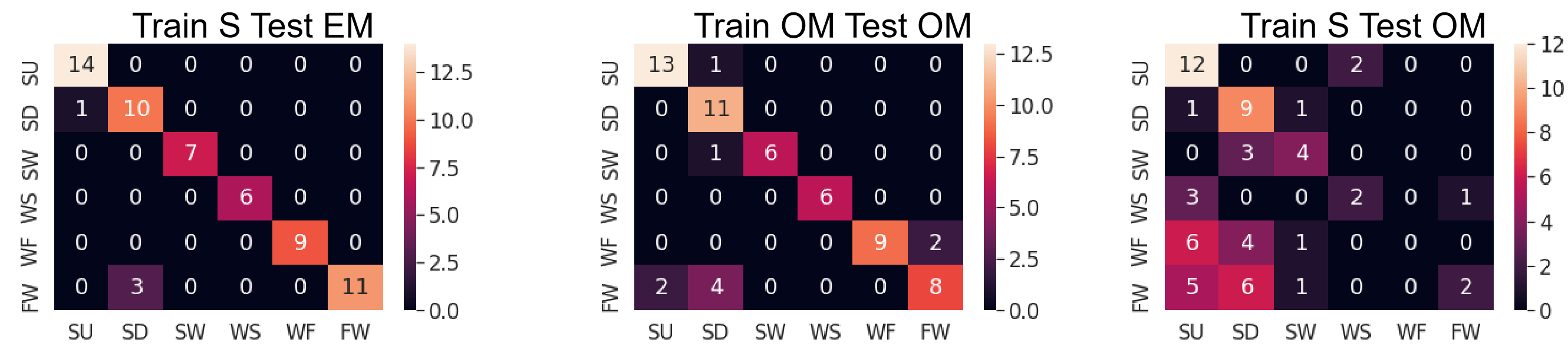}
\caption{Confusion matrices obtained from three classifiers which are trained and tested with different types of dataset: stand-up (SU), sit-down (SD), sit-to-walk (SW), walk-to-sit (WS), walk-to-fall (WF), and stand-from-floor-to-walk (FW)}
\label{fig:cms}
\end{figure}
To investigate the performance of the proposed method, we applied the VGG16 network and used same measured and simulated LOS dataset as Section \ref{subsec: dataset}. The $61$ measured spectrograms are fixed to be testing data, and the only difference relates to the use of the original measured data or the enhanced data. We respectively used only simulated data (S) and only original measured data (OM) to train classifiers, and tested them with their enhanced data (EM) and OM. As a result, we obtained three cases- 1).Train S Test EM, 2).Train S Test OM and 3).Train OM Test OM. Meanwhile, we gradually increased the number of samples in the training set to demonstrate the performance of two types of classifiers. The classification results are shown in Table \ref{tab:classification} and the confusion matrices of highest-accuracy classifiers in Train OM Test OM, Train S Test EM and Train S Test OM cases are shown in Figure \ref{fig:cms}.

We can observe that the classifier of Train S Test EM always outperforms another classifier. First of all, this indicates that the proposed classification method is feasible in the real application. Furthermore, due to the simulated data and enhanced data have better quality than the measured spectrograms, the classification accuracy can be improved with the new method. For the last row, the measured dataset cannot provide more samples to the training set, resulting in the final classification accuracy is limited to $92.1\%$. In contrast, the simulated dataset can continuously generate new samples, so the proposed method can further improve the classification accuracy from $93.7\%$ to $93.9\%$. This easily and effectively solved the issue caused by insufficient training data. Additionally, from Train S Test OM case. We can observe that the classification accuracy is far below other results. This means that the classifier can only recognize the connection between two types of data after the measured data has been enhanced with the FMNet. This once more demonstrated the FMNet enhancement framework's feasibility and superior performance.
\section{Conclusion}
\label{sec: Conclusion}
For improving the quailty of $\mu$-DS, we propose the FMNet enhancement framework. Unlike existing approaches, the method uses the feature mapping idea and bypasses multipath, clutter and interference to directly enhance motion patterns based on the similarity of latent features between measured and simulated data. The framework is based on VAE and AAE structures but assigned different purposes for each sub-networks. To progressively accomplish latent regularization, latent feature mapping, and measured spectrogram enhancement functions, we developed a three-phase training algorithm. Through qualitatively and quantitatively comparing to other networks, we demonstrated each phase is helpful and capable of improving the latent space property and algorithm robustness. In addition, we tested the FMNet trained with LOS dataset on the TTW scenario, and the proposed method still performs well. This suggests that the FMNet enhancement framework can be generalised, which is important in practical applications. Additionally, we proposed a novel classification scheme, which is a potential application of the FMNet. The comparison of results shows that our method can address the issue of a lack of training data while boosting classification accuracy with high-quality simulated and enhanced spectrograms. Based on the promising results, FMNet will continue to be investigated. 

\section*{Acknowledgments}
This work is part of the OPERA project funded by the UK Engineering and Physical Sciences Research Council (EPSRC), Grant No: EP/R018677/1. 


\bibliographystyle{IEEEtran}

\bibliography{refs.bib}

\balance

\end{document}